\documentclass[twocolumn,showpacs,preprintnumbers,amsmath,amssymb]{revtex4}
\usepackage{amsmath,amssymb,graphics,epsfig,subfigure}
\usepackage{color}

\begin{document}

\thispagestyle{empty}

\begin{center}

\title{Black Hole Solutions as Topological Thermodynamic Defects}

\date{\today}
\author{Shao-Wen Wei$^{1,2}$ \footnote{E-mail: weishw@lzu.edu.cn},
Yu-Xiao Liu$^{1,2}$ \footnote{E-mail: liuyx@lzu.edu.cn},
Robert B. Mann$^{3}$ \footnote{E-mail: rbmann@uwaterloo.ca}}

\affiliation{$^{1}$Lanzhou Center for Theoretical Physics, Key Laboratory of Theoretical Physics of Gansu Province, School of Physical Science and Technology, Lanzhou University, Lanzhou 730000, People's Republic of China,\\
$^{2}$Institute of Theoretical Physics $\&$ Research Center of Gravitation,
Lanzhou University, Lanzhou 730000, People's Republic of China,\\
$^{3}$ Department of Physics \& Astronomy, University of Waterloo, Waterloo, Ont. Canada N2L 3G1}

\begin{abstract}
In this Letter, employing the generalized off-shell free energy, we treat black hole solutions as defects in the thermodynamic parameter space. The results show that the positive and negative winding numbers corresponding to the defects indicate the local thermodynamical stable and unstable black hole solutions, respectively. The topological number defined as the sum of the winding numbers for all the black hole branches at an arbitrary given temperature is found to be a universal number independent of the black hole parameters. Moreover, this topological number only depends on the thermodynamic asymptotic behaviors of the black hole temperature at small and large black hole limits. Different black hole systems are characterized by three classes via this topological number. This number could help us in better understanding the black hole thermodynamics, and further shed new light on the fundamental nature of quantum gravity.
\end{abstract}


\maketitle
\end{center}

{\it Introduction}---
Black holes predicted by general relativity play a central role in modern physics. Observations of binary black hole mergers by the LIGO/Virgo Collaboration \cite{Abbott} and reconstruction of event-horizon-scale images of M87* by the Event Horizon Telescope (EHT) \cite{Akiyama1} have opened new windows to study the strong gravitational nature of black holes. Alternatively, seeking   underlying characteristic properties of black holes is also quite valuable, since they can yield features that can  be tested by further astronomical observations.

One such approach is topology: by ignoring  detailed structure,  generic properties of a system can be discerned.
This approach has been extensively employed to study many physical phenomena such as magnetic monopoles, the quantum Hall effect, and superconductivity. A key concept is that of defects, which are generally thought to be related to zero points of a field at $\vec{x}=\vec{z}$,
\begin{eqnarray}
 \phi(\vec{x})|_{\vec{x}=\vec{z}}=0,\label{p0}
\end{eqnarray}
in a space,  and which can reveal  certain properties of field configurations. In different dimensions, a defect can be a point charge, a string, or even a domain wall. The most simple topological quantity associated with the  zero point of a field is its winding number. When equipped with it, we can determine the nature of a system possessing defects.

Of particular recent interest has been a special vector   constructed in the coordinate space of a black hole spacetime \cite{Cunhaa,Cunhab} by making use of  null geodesics. The light ring of a black hole is located exactly at the zero point of this vector field. After calculating the winding number corresponding to this zero point,  a four-dimensional stationary nonextremal spinning black hole in asymptotically flat spacetime with a topologically spherical Killing horizon was shown to allow at least one standard light ring outside the horizon for each sense of rotation \cite{Cunhab}. This treatment was generalized to other cases, where the critical points and timelike circular orbit were well studied \cite{Wei2019,Guo,Wei2112,Ahmed,Yerra,Bhamidipati,Wei2022}.

In general relativity, a black hole solution satisfies the Einstein field equations, which we reformulate as
\begin{eqnarray}
\mathcal{E}_{\mu\nu} \equiv G_{\mu\nu}-\frac{8\pi G}{c^4} T_{\mu\nu}=0.\label{pp}
\end{eqnarray}
We propose, analogous to  Eq.~(\ref{p0}),  that a physical black hole solution is a zero point of the tensor field $\mathcal{E}_{\mu\nu}$, at which all its components vanish. Although the Einstein field equations also admit other defect-like solutions, such as cosmic strings and branes, we here only focus on  black holes. As a result, we can endow a black hole solution with a topological charge as well. Adopting this topological argument, we can then study the local and global particular properties of a black hole. Different black hole solutions are also expected to be divided into different classes.

{\it Realization in thermodynamics}--- As a starting point to realize the idea that a black hole solution possesses a topological charge, we begin with black hole thermodynamics.

In 1977, Gibbons and Hawking \cite{Gibbons} proposed that the partition function of a canonical ensemble for black holes can be evaluated with its Euclidean action in the form of the gravitational path integral. In ``zero-loop" approximation, it reads
\begin{eqnarray}
 \mathcal{Z}=e^{-\beta F}=\int D[g]e^{-\frac{\mathcal{I}}{\hbar}}\sim e^{-\frac{\mathcal{I}}{\hbar}},\label{actionz}
\end{eqnarray}
where $F$ and $\mathcal{I}$ is the free energy and Euclidean action of the black hole. The period $\beta$ of the Euclidean time is the inverse of the black hole temperature. Following this approach, it was found that negative heat capacity and imaginary energy fluctuations were produced. Soon afterward, these shortcomings were solved by York \cite{York} by imagining that the black hole is placed inside a cavity. Fixing the temperature $T$ of the cavity surface, the results show that the heavy-mass black hole branch with mass $M>\sqrt{3}/8\pi T$ is thermodynamically stable, and thus the corresponding partition function is well defined.

In order to describe a black hole with an arbitrary mass, York defined a generalized free energy following the standard definition of the free energy with the mass and temperature being two independent variables \cite{York}. This naturally extends one more dimension for the thermodynamic parameter space of the black hole. Furthermore,    the generalized free energy   reduces  to the standard one when the relation between the black hole mass and temperature is satisfied \cite{Whiting}.

Inspired by this, we would like to introduce the generalized free energy
\begin{eqnarray}
 \mathcal{F}=E-\frac{S}{\tau},\label{grf}
\end{eqnarray}
for a black hole system with energy $E$ and entropy $S$. The parameter $\tau$ is an extra variable having the dimension of time, and can be thought as the inverse temperature of the cavity enclosing the black hole. Here we let the time parameter $\tau$ vary freely instead of the mass \cite{York}. In general this generalized free energy is off-shell except
at
\begin{eqnarray}
 \tau=T^{-1},\label{condd}
\end{eqnarray}
at which the black hole solution satisfies the Einstein field equations (\ref{pp}). Recently, a similar generalized free energy (\ref{grf}) has been used to study the dynamic evolution of black hole phase transitions \cite{Li,Wang,weiMann,Mann}.

We now describe the explicit construction for the vector $\phi$ in~(\ref{p0}) via a thermodynamic approach. We define
\begin{eqnarray}
 \phi=\left(\frac{\partial \mathcal{F}}{\partial r_{\text{h}}}, -\cot\Theta \csc\Theta\right) \label{vectorField}
\end{eqnarray}
where, inspired by the axis limit \cite{Cunhab}, we introduce a parameter $0\leq\Theta\leq\pi$ for convenience. At $\Theta=0$, $\pi$, the component $\phi^{\Theta}$ diverges and thus the direction of the vector is outward.

More importantly, the zero points of $\phi$ correspond to the conditions $\Theta=\pi/2$ and $\tau=T^{-1}$. This confirms that the black hole solution exactly meets the zero point of the vector $\phi$. Therefore, from the viewpoint of topology, we can endow each black hole solution with a topological charge by using $\phi$.

Following Duan's $\phi$-mapping topological current theory \cite{Duana,Duanb}, we can introduce the topological current as
\begin{eqnarray}
 j^{\mu}=\frac{1}{2\pi}\epsilon^{\mu\nu\rho}\epsilon_{ab}\partial_{\nu}n^{a}\partial_{\rho}n^{b},
 \quad \mu,\nu,\rho=0,1,2,
\end{eqnarray}
where $\partial_{\nu}=\frac{\partial}{\partial x^{\nu}}$ and $x^{\nu}=(\tau,~r_{\text{h}},~\Theta)$. The unit vector is defined as $n^a=\frac{\phi^a}{||\phi||}$ ($a=1, 2$) with $\phi^1=\phi^{r_{\text{h}}}$ and $\phi^2=\phi^\Theta$. Here the parameter $\tau$ can be considered as a time parameter of the topological defect. Moreover, it is easy to check that this topological current is conserved, i.e., $\partial_{\mu}j^{\mu}=0$. By making use of the Jacobi tensor $\epsilon^{ab}J^{\mu}\left(\frac{\phi}{x}\right)=\epsilon^{\mu\nu\rho}
\partial_{\nu}\phi^a\partial_{\rho}\phi^b$ and the two-dimensional Laplacian Green function $\Delta_{\phi^a}\ln||\phi||=2\pi\delta^2(\phi)$, the topological current can be reexpressed as \cite{Wei2019}
\begin{equation}
 j^{\mu}=\delta^{2}(\phi)J^{\mu}\left(\frac{\phi}{x}\right).\label{juu}
\end{equation}
Obviously, $j^{\mu}$ is nonzero only at $\phi^a(x^i)=0$, and we denote its $i$-th solution as $\vec{x}=\vec{z}_{i}$.
The density of the topological current is then \cite{Schouton}
\begin{equation}
 j^{0}=\sum_{i=1}^{N}\beta_{i}\eta_{i}\delta^{2}(\vec{x}-\vec{z}_{i}).
\end{equation}
The positive Hopf index $\beta_i$ counts the number of the loops that $\phi^a$ makes in the vector $\phi$ space when $x^{\mu}$ goes around the zero point $z_i$, and the Brouwer degree $\eta_{i}=\text{sign}(J^{0}({\phi}/{x})_{z_i})=\pm1$. Given a parameter region $\Sigma$, the corresponding topological number can be obtained
\begin{eqnarray}
 W=\int_{\Sigma}j^{0}d^2x
 =\sum_{i=1}^{N}\beta_{i}\eta_{i}=\sum_{i=1}^{N}w_{i},
\end{eqnarray}
where $w_{i}$ is the winding number for the $i$-th zero point of $\phi$ contained in $\Sigma$. If two given loops $\partial\Sigma$ and $\partial\Sigma'$ enclose the same zero point of $\phi$, they possess the same winding number. Alternatively,  if there is no zero point  in the enclosed region, we will have $W=0$. If $\Sigma$ is the neighbourhood of a zero point of $\phi$, it will yield  local topological properties, whereas if $\Sigma$  is the entire parameter space, the global topological $W$ number will be revealed.

In the above approach, the isolated zero points require the Jacobian $J_{0}(\phi/x)\neq0$. If this condition is violated,  the defect bifurcates \cite{Fu}.

{\it Local and global topological properties}---Following Eq.~(\ref{actionz}), one can obtain the free energy by calculating the first-order Euclidean Einstein action
\begin{eqnarray}
 \mathcal{I}=-\frac{1}{16\pi}\int\sqrt{g}Rd^4x+\frac{1}{8\pi}\oint\sqrt{\gamma}Kd^3x-\mathcal{I}_{subtract}
\end{eqnarray}
where $K$ is the trace of the extrinsic curvature, $\gamma_{ij}$ is the induced metric on the boundary, and $\mathcal{I}_{subtract}$ is the subtraction term.

\begin{figure}
\subfigure[]{\label{Topo}\includegraphics[width=5cm]{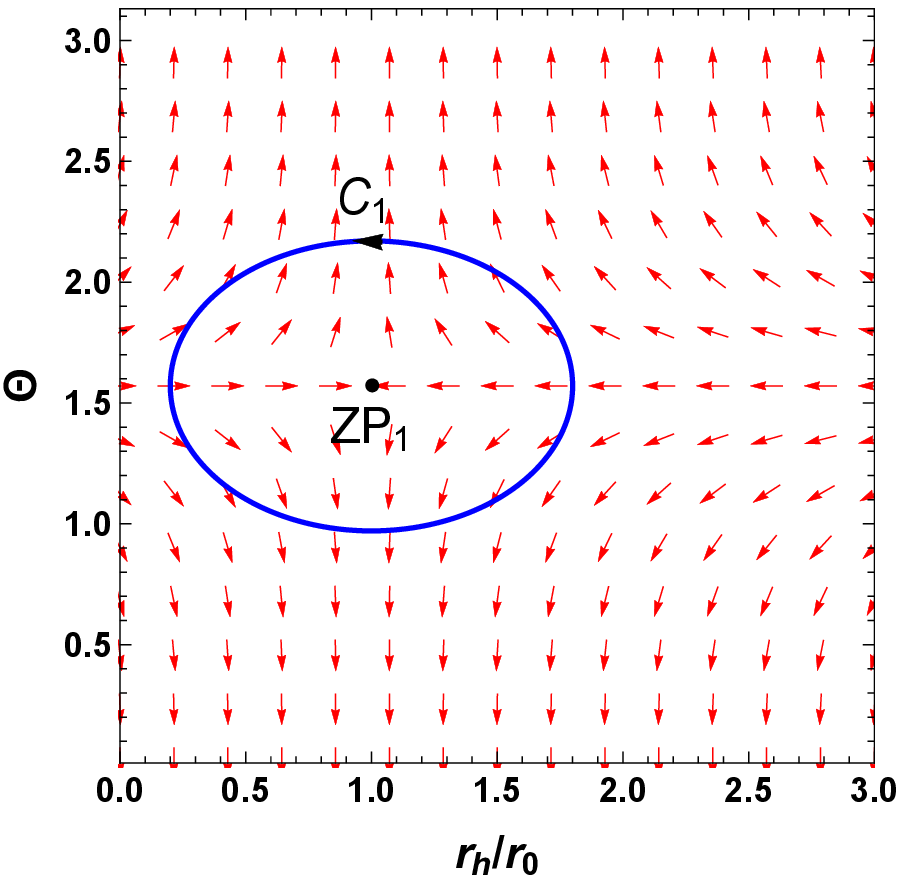}}
\subfigure[]{\label{TopoRN}\includegraphics[width=5cm]{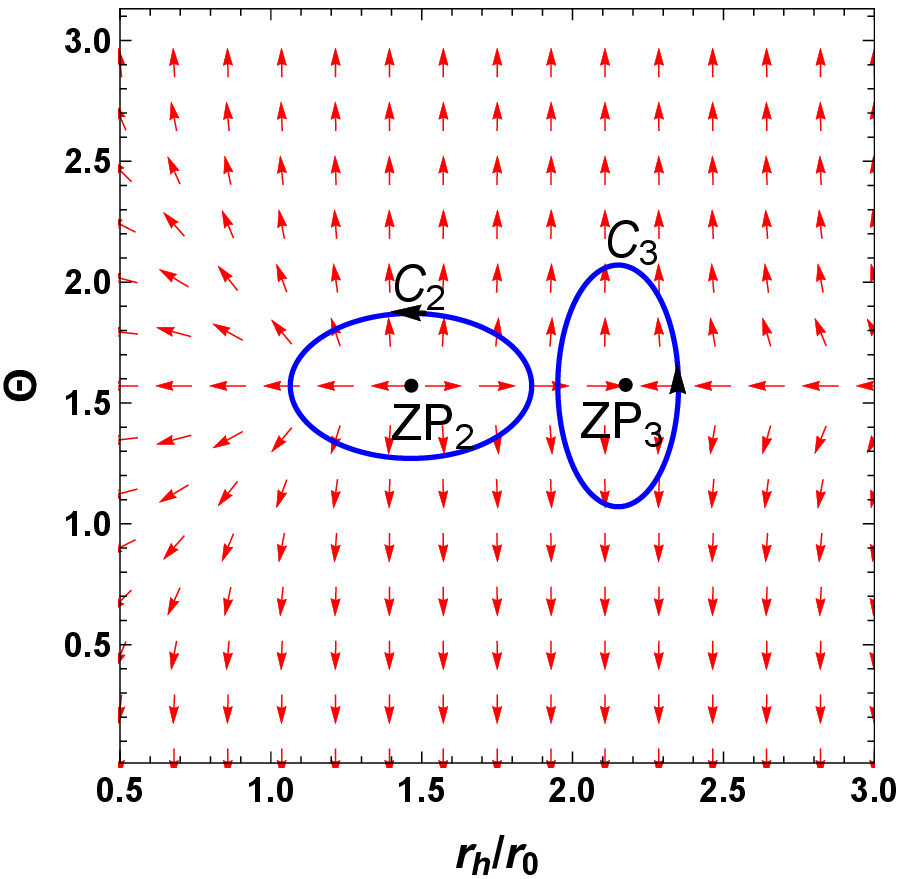}}
\caption{The red arrows represent the unit vector field $n$ on a
portion of the $r_{\text{h}}$-$\Theta$ plane. The zero points (ZPs) marked with black dots are at ($r_{\text{h}}$, $\Theta$)=(1, $\frac{\pi}{2}$), (1.46, $\frac{\pi}{2}$), and (2.15, $\frac{\pi}{2}$), for ZP$_1$, ZP$_2$, and ZP$_3$, respectively. The blue contours $C_i$ are closed loops enclosing the zero points. (a) The unit vector field for the Schwarzschild black hole with $\tau/r_0=4\pi$. (b) The unit vector field for the RN black hole with $\tau/r_0=34.48$ and $Q/r_0=1$.}\label{pCharVec}
\end{figure}

For the Schwarzschild black hole, one easily has $E=\partial_\beta\mathcal{I}=M$ and $S=
(\beta \partial_\beta\mathcal{I}-\mathcal{I})=4\pi M^2$ \cite{Gibbons}. Accordingly, the black hole thermodynamic relations can be deduced straightforwardly. For the Schwarzschild black hole, we obtain the generalized free energy $\mathcal{F}=\frac{r_{\text{h}}}{2}-\frac{\pi r_{\text{h}}^2}{\tau}$, where $r_{\text{h}}=2M$. Then the components of the constructed vector $\phi$ can be calculated
\begin{eqnarray}
 \phi^{r_{\text{h}}}&=&\frac{1}{2}-\frac{2\pi r_{\text{h}}}{\tau},\\
 \phi^{\Theta}&=&-\cot\Theta \csc\Theta.
\end{eqnarray}
We show the unit vector field $n$ on a portion of the $\Theta$-$r_{\text{h}}$ plane in Fig.~\ref{Topo} for the Schwarzschild black hole with $\tau=4\pi r_0$ with $r_0$ an arbitrary length scale set by the size of a cavity surrounding the black hole.

From the figure, the zero point is   located at $r_{\text{h}}/r_0=1$  and $\Theta=\pi/2$. Since the winding number $w$ is independent of these loops enclosing the zero point, we can calculate it for an arbitrary loop; for example see $C_1$ given in Fig. \ref{Topo}. Performing the calculation, we obtain the winding number $w=-1$.  If we choose  an alternate orientation convention by adding a minus sign in $\phi^{r_{\text{h}}}$ in (\ref{vectorField}), we obtain $w=1$ instead. However the winding numbers of other types of black holes will likewise be changed.

Considering the gravitational action together with the electromagnetic field, it is easy to find $E=\frac{r_{\text{h}}^2+Q^2}{2 r_{\text{h}}}$ and $S=\pi r_{\text{h}}^2$ for the Reissner-Nordstr\"{o}m (RN) black hole. Thus the generalized free energy is
\begin{eqnarray}
 \mathcal{F}=\frac{r_{\text{h}}^2+Q^2}{2 r_{\text{h}}}-\frac{\pi r_{\text{h}}^2}{\tau}.
\end{eqnarray}
Here $\tau$ can be thought as the inverse temperature of the cavity. Thus $r_{\text{h}}$ is independent of $\tau$, and can vary freely, ignoring the lower bound. Employing it, we plot the unit vector field $n$ in Fig. \ref{TopoRN} for  arbitrarily chosen typical values $\tau/r_0=34.48$ and $Q/r_0=1$.

We find two zero points, ZP$_2$ and ZP$_3$,  at $r_{\text{h}}/r_0=1.46$ and $2.15$ respectively. A detailed calculation shows that their respective winding numbers are $w=1$ and $-1$.  We also obtain the   heat capacities  $C_{Q}/r_{0}^2=18.01$ and $-64.81$ at constant charge $Q/r_0=1$ for the respective positive and negative zero points.  Noting that each zero point of the unit vector has a winding number $1$ or $-1$, from this we conjecture that  winding number is related to  local thermodynamic stability, with  positive/negative values corresponding to stable/unstable black hole solutions.

Turning  to global properties of the topology, if we choose the region $\Sigma$ as the whole parameter space or the loop $\partial\Sigma$ as the boundary of the parameter space, i.e., $(0<\Theta<\pi)\cup(0<r_{\text{h}}<\infty)$, we can obtain the topological number $W$ for a black hole solution. This   global property    can be used to classify   different black hole solutions. Since we here take the axis limit, where the direction of the vector $\phi$
is up at $\Theta=\pi$ and down at $\Theta=0$, the value of the topological number $W$ actually depends on the direction of the vector at $r_{\text{h}}=0$ and $\infty$. This suggests that black hole solutions sharing the same behavior at $r_{\text{h}}=0$ and $\infty$ possess the same topological number, and thus  are topologically equivalent.

\begin{figure}
\includegraphics[width=7cm]{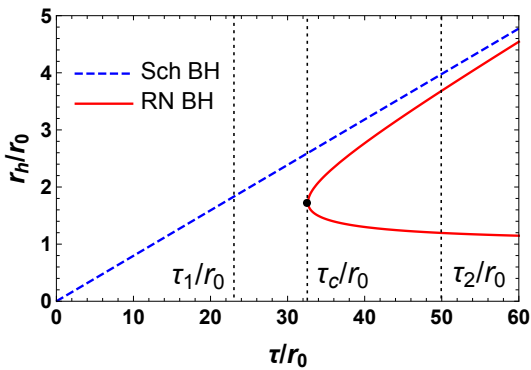}
\caption{Zero points of the vector $\phi$ shown in the $r_{\text{h}}$-$\tau$ plane. The blue dashed and red solid lines are for the Schwarzschild black hole (Sch BH) and RN black hole (BH) with $Q/r_0=1$. The black dot with $\tau_c=6\sqrt{3}\pi Q$ denotes the generation point for the RN black hole. At $\tau=\tau_1$, there is one Schwarzschild black hole, and at $\tau=\tau_2$, there are one Schwarzschild black hole and two RN black holes.}\label{rht}
\end{figure}

By solving the equation $ \phi^{r_{\text{h}}}=0$, we can obtain a curve in the $r_{\text{h}}-\tau$ plane. The results are
\begin{eqnarray}
\tau=4\pi r_{\text{h}}  \label{Curvetaurh_Sch}
\end{eqnarray}
and
\begin{eqnarray}
 \tau=\frac{4\pi r_{\text{h}}^3}{r_{\text{h}}^2-Q^2}  \label{Curvetaurh_RN}
\end{eqnarray}
for the Schwarzschild and RN black holes, respectively.
To show zero points of the component $ \phi^{r_{\text{h}}}$, we plot these two curves in Fig. \ref{rht}. For large $\tau$ (e.g. $\tau=\tau_2$) there are respectively  one and two intersection points for the Schwarzschild and RN black holes. The intersection points exactly satisfy the condition (\ref{condd}), and therefore  denote the on-shell black hole solutions with temperature $T=\tau^{-1}$. In contrast to the Schwarzschild black hole, for $\tau < \tau_{c}$, the two intersection points for the RN black hole coincide and then disappear. Based on the local property of a zero point, we have the topological number $W=-1$ for the Schwarzschild black hole, while $W=1-1=0$ for the RN black hole with the charge $Q/r_0=1$.
In particular, at the point $\tau_{c}=6\sqrt{3}\pi Q$, it is easy to get $\frac{d^2\tau}{dr_{\text{h}}^2}=6\sqrt{3}\pi/Q>0$ for the RN black hole. This suggests that $\tau_{c}$  is a generation point, which can also be observed in Fig. \ref{rht}.

To investigate whether or not $W$ depends on the charge, we examine the behavior of the curve $\tau(r_{\text{h}})$ in Eq.~(\ref{Curvetaurh_RN}) at the limit $r_{\text{h}}\rightarrow r_{\text{E}}=Q$ (corresponding to extremal RN black hole with smallest horizon) and $\infty$. Obviously, for a nonvanishing charge $Q$, the vanishing/diverging behaviour of $\tau$  as $r_{\text{h}}\rightarrow r_{\text{E}}/\infty$ does not change, and so  $W$ remains the same for different values of $Q$. For the Schwarzschild black hole with $Q=0$, the behaviour of $\tau(r_{\text{h}})$ for small $r_{\text{h}}$ differs from the charged case, and so  the topological numbers $W$ for the two black hole solutions are different.

\begin{figure}
\includegraphics[width=7cm]{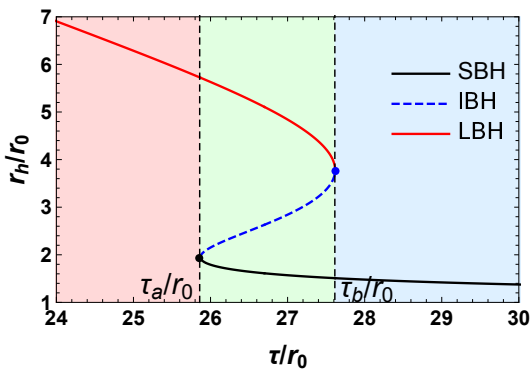}
\caption{Zero points of $\phi^{r_{\text{h}}}$ shown in the $r_{\text{h}}$-$\tau$ plane for the RN-AdS black hole with $Pr_{0}^{2}=0.0022$ and $Q/r_0=1$. The black solid, blue dashed, and red solid lines are for the small black hole (SBH), intermediate black hole (IBH), and large black hole (LBH), respectively. Black and blue dots are the annihilation and generation points. Different color regions have different numbers of the black hole branches. However their $W$ number is constant and equals 1.}\label{rhAdS}
\end{figure}

Another interesting black hole solution is the charged Reissner-Nordstr\"{o}m anti-de Sitter (RN-AdS) black hole, which exhibits a small-large black hole phase transition \cite{Kubiznak}. In the extended phase space, the cosmological constant is treated as the pressure $P$ of the system \cite{Kastor}. The free energy is
\begin{eqnarray}
 \mathcal{F}=\frac{8 \pi  P r_{\text{h}}^4+3 r_{\text{h}}^2+3 Q^2}{6 r_{\text{h}}}-\frac{\pi r_{\text{h}}^2}{\tau}.
\end{eqnarray}
Taking the pressure $Pr_{0}^{2}=0.0022$ to be smaller than its critical value, we exhibit zero points of $\phi^{r_{\text{h}}}$ in the $r_{\text{h}}-\tau$ plane in Fig. \ref{rhAdS} with $Q/r_0=1$. Quite different from the Schwarzschild and RN black holes shown in Fig. \ref{rht}, we observe that there are three black hole branches for $\tau_{a}<\tau<\tau_{b}$, one small black hole branch for $\tau<\tau_{a}$  and one large black hole branch for $\tau>\tau_{b}$. Computing the winding number for these black hole branches, we find that the small and large black hole branches have $w=1$, while the intermediate black hole branch has $w=-1$. Therefore, at this pressure the RN-AdS black hole always has $W=1-1+1=1$, which is independent of $\tau$. Since the pressure $P$ is positive for the RN-AdS black hole, it does not affect the asymptotic behavior of $\tau$ at small and large $r_{\text{h}}$. As a result, the topological number $W$ is always unity for the RN-AdS black hole. Furthermore, our result that thermodynamically stable small and large black holes have $w=1$ and unstable intermediate black holes have $w=-1$ supports our conjecture that positive/negative winding number is indicative of  thermodynamic stablilty/instability.

Note that for these values of $Pr_{0}^{2}$ and $Q/r_0$, one generation point and one annihilation point can be found at $\tau/r_0=\tau_a/r_0=25.84$ and $\tau/r_0=\tau_b/r_0=27.62$, respectively. If the pressure is larger than its critical value, $r_{\text{h}}/r_0$ will be a monotonically increasing function of $\tau/r_0$, and thus no bifurcation phenomenon will be observed. However the topological number $W$ still stays the same.

\begin{table}[h]
\setlength{\tabcolsep}{3mm}{
\begin{center}
\begin{tabular}{cccc}
  \hline\hline
 &Sch BH & RN BH &RN-AdS BH\\\hline
$W$&-1&0&1\\
generation point&0&1&1 or 0\\
annihilation point&0&0&1 or 0\\
\hline\hline
\end{tabular}
\caption{The topological number $W$, numbers of annihilation and generation points for the Schwarzschild, RN, and RN-AdS black holes.}\label{tab1}
\end{center}}
\end{table}

In the Supplemental Material we show that $d$-dimensional RN-AdS black holes also have  topological number $W=1$, consistent with the $d=4$ case (see Supplemental Material).

{\it Conclusions}---
Summarizing our results in Table \ref{tab1}, we have constructed a universal topological number $W$ by treating the black hole solution as a defect from the viewpoint of thermodynamics. Different black hole solutions are characterized by different topological numbers and belong to different topological classes.

Locally, each zero point of the vector field defined in~(\ref{vectorField}) with the generalized free energy exactly corresponds to one on-shell black hole solution. Thus each black hole solution is endowed with one winding number. Our study shows that the thermodynamic stability of the black hole is indicated from the value of the winding number. Positive/negative winding number corresponds to  thermodynamically stable/unstable black hole solutions. Of particular interest are the annihilation and generation points, which may be quite important for the time evolution of a black hole placed in a cavity.

When considering the full parameter space, the topological number $W$ provides us with a global topological property of a black hole solution. We have shown that the Schwarzschild black hole, RN black hole, and RN-AdS black hole have respectively $W=-1,~0$, and $1$, independent of the black hole parameters. Since the topological number $W$ is the sum of the winding numbers for the zero points of the vector, and it is only dependent on the behavior of the curve $\tau(r_{\text{h}})$ at small and large $r_{\text{h}}$ limits, we conjectured that the topological number $W$ can take three values, $-1$, $0$, and $1$. This suggests that other black hole solutions shall be divided into three characteristic topological classes, a conjecture in
need of  further confirmation.

We close by considering whether geometric modifications to black holes, such as those induced by scalar hair,  affect the topological number $W$. In contrast to global charges (primary hair), black holes possessing new non-trivial fields (secondary/pseudo hair) have long been of interest in string theory  \cite{Campbell,Kaloper,Kanti} and have attracted much recent attention  in both GR and modified gravity theories \cite{Herdeiro,Herdeirorev,Silva,Radu,Cunha,Collodel,Zhang,Konoplya}. For the Einstein-Maxwell-scalar model \cite{Radu,Konoplya}, we show in the Supplemental Material that the topological number $W=0$ for these scalarized black holes, the same as that of the RN black holes, and in support of the expectation that secondary (scalar) hair does not change $W$ despite the fact that it modifies the free energy of a black hole (see Supplemental Material). In more general string-theoretic settings  (for example axion fields with Lorentz Chern-Simons coupling to gravity \cite{Campbell}) the generalized free energy will be modified.  Investigation of the topological charge for these more general black holes with hair is an interesting project for further study.

In conclusion, our topological approach classifies each black hole solution into certain classes sharing the similar thermodynamic properties. It provides a considerable material for the topology of black hole thermodynamics. We expect to uncover the deeper nature of other black hole solutions via the topological approach.

{\emph{Acknowledgements}.}---This work was supported by the National Natural Science Foundation of China (Grants No. 12075103, No. 11875151, and No. 12047501), the 111 Project (Grant No. B20063), and the National Key Research and Development Program of China (Grant No. 2020YFC2201503) and by the Natural Sciences and Engineering Research Council of Canada.

{\it \textbf{Supplemental Material A: Higher dimensional black holes}}

The bulk action describing the $d$-dimensional charged Reissner-Nordstr\"{o}m (RN)  AdS black hole solution reads
\begin{eqnarray}
  \mathcal{I}=-\frac{1}{16}\int d^{d}x\sqrt{-g}\left(R-F^2+\frac{(d-1)(d-2)}{l^2}\right).
\end{eqnarray}
Solving the resultant Einstein-Maxwell equations we obtain the black hole metric
\begin{eqnarray}
  ds^2=-fdr^2+f^{-1}dr^2+r^2 d\Omega^{2}_{d-2},\\
  F=dA,\quad
  A=-\sqrt{\frac{d-2}{2(d-3)}}\frac{q}{r^{d-3}}dt
\end{eqnarray}
where
\begin{eqnarray}\label{bh1}
  f=1-\frac{m}{r^{d-3}}+\frac{q^2}{r^{2(d-3)}}+\frac{r^2}{l^2}.
\end{eqnarray}
The Arnowitt-Deser-Misner mass $M$ and charge $Q$ of the black hole, respectively, are
\begin{eqnarray}
  M=\frac{d-2}{16\pi}\omega_{d-2}m, \quad
  Q=\frac{\sqrt{2(d-2)(d-3)}}{8\pi}\omega_{d-2}q,
\end{eqnarray}
with $\omega_{d-2}=2\pi^{(d-1)/2}/\Gamma((d-1)/2)$ being the volume of the unit ($d-2$)-sphere. In the extended phase space, one has the thermodynamic pressure $P=\frac{(d-1)(d-2)}{16\pi l^2}$. For fixed charge  the total action can be calculated by adding a boundary term \cite{Chamblin}
\begin{eqnarray}
  \mathcal{\tilde{I}}=\mathcal{I}-\frac{1}{4\pi}\int d^{d-1}x\sqrt{h}F^{\mu\nu}n_{\mu}A_{\nu},
\end{eqnarray}
where $n_{\mu}$ is a radial unit vector pointing outwards.

For the black hole solution \eqref{bh1}, a computation of the action yields  the free energy $F=\mathcal{\tilde{I}}/\beta=M-TS$.
Using the topological approach we proposed, we obtain
\begin{eqnarray}
  \tau=\frac{4 \pi  (d-2) \omega_{d-2}^2 r_{\text{h}}^{2
   d+1}}{\omega_{d-2}^2 r_{\text{h}}^{2 d} \left((d-5)
   d+16 \pi  P r_{\text{h}}^2+6\right)-32 \pi ^2 Q^2
   r_{\text{h}}^6}
\end{eqnarray}
as the zero point of the vector field.

\begin{figure}
\includegraphics[width=7cm]{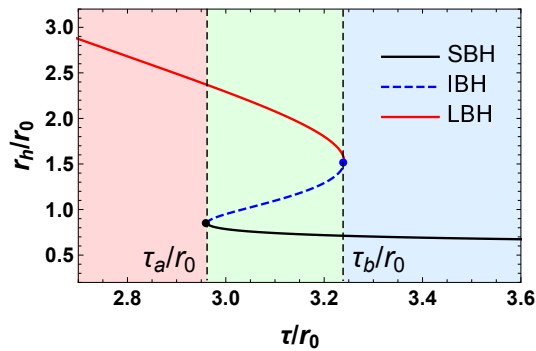}
\caption{Zero points of $\phi^{r_{\text{h}}}$ shown in the $r_{\text{h}}$-$\tau$ plane for the $d=6$-dimensional RN-AdS black hole with $Pr_{0}^{2}=0.1$ and $Q/r_{0}^{3}=1$. The black solid, blue dashed, and red solid lines are for the small, intermediate, and large black hole branches, respectively. This pattern is quite similar to the $d=4$-dimensional case.}\label{d6rht}
\end{figure}

As a specific example, we take $d=6$, $Q/r_{0}^3=1$, and $Pr_{0}=0.1$, and show the zero points of $\phi$ in the $r_{\text{h}}$-$\tau$ plane in Fig. \ref{d6rht}. Obviously, the behavior is similar to the four-dimensional black hole case exhibited in Fig. 3 in the main text. The result indicates that the topological number $W=1$ for the six-dimensional RN-AdS black hole. In order to examine whether the topological number
has any dimension dependence, we respectively expand $\tau$ at small and large $r_{\text{h}}$, obtaining
\begin{eqnarray}
  &&\tau\sim\infty \quad\text{for}\quad r_{\text{h}}\rightarrow  r_{\text{E}},\\
  &&\tau\sim\frac{(d-2)\pi}{4\pi P r_{\text{h}}} \quad\text{for}\quad r_{\text{h}}\rightarrow \infty,
\end{eqnarray}
where $r_{\text{E}}$ denotes the horizon radius of the extremal black hole with zero temperature and qualtitatively the same as Eqs. (15) and (16) in the main text. Hence the topological number is the same as the $d=4$ RN-AdS black hole, and the spacetime dimension  $d$  thus has no effect on the topological number.   We conclude that all   RN-AdS black holes have $W=1$, and  belong to the same topological class.

Furthermore, in any dimension  small and large charged AdS black holes are thermodynamical stable, whereas their intermediate counterparts are unstable. In order to obtain $W=1$,  the small and large black hole must have winding number $+1$ and intermediate one has $-1$, which uniquely gives $W=1+1-1=1$ as expected. A simple calculation will also show the same result. This further indicates that thermodynamical stable or unstable black hole has positive or negative winding number.

{\it \textbf{Supplemental Material B: Scalarized black holes}}

Here we investigate the effect of black hole scalar hair on the topological number $W$.

To this end we consider the following Einstein-Maxwell-scalar model  \cite{Radu,Konoplya}
\begin{eqnarray}
  S=\int d^4\sqrt{-g}\left(R-2g^{\mu\nu}\partial_{\mu}\phi\partial_{\nu}\phi-f(\phi)F_{\mu\nu}F^{\mu\nu}\right),
  \label{aact}
\end{eqnarray}
whose field equations yield different spherically symmetric scalarized black holes for
 different scalar functions $f(\phi)$. For the ansatz
\begin{eqnarray}
  ds^2=-N(r)e^{-2\delta(r)}dt^2+\frac{dr^2}{N(r)}+r^2d\Omega^{2}_{2},
\end{eqnarray}
we must have
\begin{eqnarray}
  \lim_{r\rightarrow\infty}N(r)=1,\quad  \lim_{r\rightarrow\infty}\delta(r)=0,
  \label{rhrh}
\end{eqnarray}
assuming that the metic is asymptotically  Minkowskian and that time is measured by the coordinate $t$.
We further assume that such scalarized black holes are of the RN type, meaning that there exists a lower bound of the horizon radius, at which the temperature of the black hole vanishes.

In order to evaluate the free energy of the black hole, we substitute the back hole solution for the action (\ref{aact}), including both  boundary and subtraction terms. It is clear that the free energy is modified by the scalar, and so scalarized black holes exhibit thermodynamic behaviour different from their RN counterparts.

Despite this, we can provide some results about the topological number for these scalarized black holes. The zero point of the vector constructed  in our topological approach can be written as
\begin{eqnarray}
  \tau=\frac{4\pi r_{\text{h}}e^{\delta(r_{\text{h}})}}{1-\frac{Q^2}{f(p_0)r_{\text{h}}^2}},
\end{eqnarray}
where $Q$ is the black hole charge and $p_0>0$ is the value of scalar field at the horizon. The denominator of $\tau$ is nonnegative, and vanishes for the extremal scalarized black hole with vanishing temperature. So similar to the RN black hole, near the lower bound of the horizon radius, we have
\begin{eqnarray}
  \tau\sim \infty \quad\text{for}\quad r_{\text{h}}\rightarrow r_{\text{E}}.
\end{eqnarray}
Making use of (\ref{rhrh}), one knows $e^{\delta(r_{\text{h}})}\rightarrow1$ for large $r_{\text{h}}$. Therefore, we have
\begin{eqnarray}
  \tau\sim 4\pi r_{\text{h}} \quad\text{for}\quad r_{\text{h}}\rightarrow \infty.
\end{eqnarray}
This result is exactly consistent with  Eqs. (15) and (16) in the main text for the RN black hole. Hence the topological charge $W=0$ for this class of scalarized black holes. Of notable  interest is that the scalar function $f(\phi)$ makes no contribution to the asymptotic behavior of $\tau$.

Summarizing, in this Einstein-Maxwell-scalar model, the scalar field does not change the topological charge $W$. However, the existence of more than one pair stable and unstable scalarized black hole branches remains a possibility, and so the sum of the local winding numbers $w$ could nontrivially yield $W=0$.

\end{document}